\documentclass[prl,twocolumn,superscriptaddress,aps]{revtex4-1}
\usepackage{amsmath}
\usepackage{amsfonts}
\usepackage{amssymb}
\usepackage{color}
\usepackage[normalem]{ulem}
\usepackage{dsfont}
\usepackage{dcolumn}
\usepackage{bm}
\usepackage{slashed}
\usepackage[toc,page]{appendix}
\usepackage{float}
\usepackage{graphicx}

\begin{document}
\bibliographystyle{apsrev4-1}
%

\title{Robust Transport Signatures of Topological Superconductivity in Topological Insulator
Nanowires}

\author{Fernando de Juan}
\affiliation{Materials Science Division, Lawrence Berkeley National Laboratory, Berkeley, CA
94720}
\affiliation{Department of Physics, University of California, Berkeley, CA 94720, USA}
\author{Roni Ilan}
\affiliation{Department of Physics, University of California, Berkeley, CA 94720, USA}
\author{Jens H. Bardarson}
\affiliation{Department of Physics, University of California, Berkeley, CA 94720, USA}
\affiliation{Max-Planck-Institut f\"ur Physik komplexer Systeme, N\"othnitzer Str.\ 38, 01187
Dresden, Germany}
\date{\today}

\begin{abstract}
Finding a clear signature of topological superconductivity in transport experiments remains an
outstanding challenge. In this work, we propose exploiting the unique properties of
three-dimensional topological insulator nanowires to generate a normal-superconductor junction
in the single-mode regime where an exactly quantized $2e^2/h$ zero-bias conductance can be observed
over a wide range of realistic system parameters. This is achieved by inducing superconductivity in
half of the wire, which can be tuned at will from trivial to topological with a parallel magnetic
field, while a perpendicular field is used to gap out the normal part, except for two spatially
separated chiral channels. The combination of chiral mode transport and perfect Andreev reflection
makes the measurement robust to moderate disorder, and the quantization of conductance survives to
much higher temperatures than in tunnel junction experiments. Our proposal may be understood as a
variant of a Majorana interferometer which is easily realizable in experiments. 
\end{abstract}




\maketitle

A topological superconductor is a proposed novel phase of matter with exotic properties like
protected boundary states and emergent quasiparticles with non-Abelian statistics. If realized,
these superconductors are expected to constitute the main building block of topological quantum
computers~\cite{NSS08}. The prototypical example of this phase, the $p$-wave superconductor, has
proven to be difficult to find in nature, with superconducting Sr$_2$RuO$_4$ and, indirectly, the
$\nu=\text{5/2}$ fractional quantum Hall state among the very few conjectured candidates. While
many experiments have been suggested and performed on these systems, evidence
for their topological properties remains elusive. However, the recent realization that a $p$-wave
superconductor need not be intrinsic, but can alternatively be engineered with regular $s$-wave
superconducting proximity effect in strongly spin-orbit coupled materials \cite{FK08,A12,B13}, has
opened a promising new path in the search for topological superconductivity. 

A class of these new topological superconductors is predicted to be realized in
one-dimensional (1D) systems with broken time-reversal symmetry~\cite{K01}. These
systems are characterized by Majorana zero-energy end states, which are responsible for a
fundamental transport effect known as perfect (or resonant) Andreev
reflection~\cite{LLN09}: in a junction between a normal contact that hosts a
single propagating mode and a topological superconductor, this mode must be perfectly reflected
as a hole with unit probability, resulting in the transfer of a Cooper pair across the junction and
an \emph{exactly} quantized zero-bias conductance of $2e^2/h$. This effect does not depend at all on
the details of the junction, and can be intuitively understood as resonant transport mediated by
the Majorana end states~\cite{LLN09,A12}. On the other hand, if the superconductor is trivial and
hence has no Majorana state, in the single-mode regime the conductance exactly vanishes. The
conductance in the single-mode regime is in fact a topological invariant~\cite{FHA11,WAD11} that
directly distinguishes trivial from topological superconductors in a transport experiment.


\begin{figure}[t]
\begin{center}
\includegraphics[width=8.2cm]{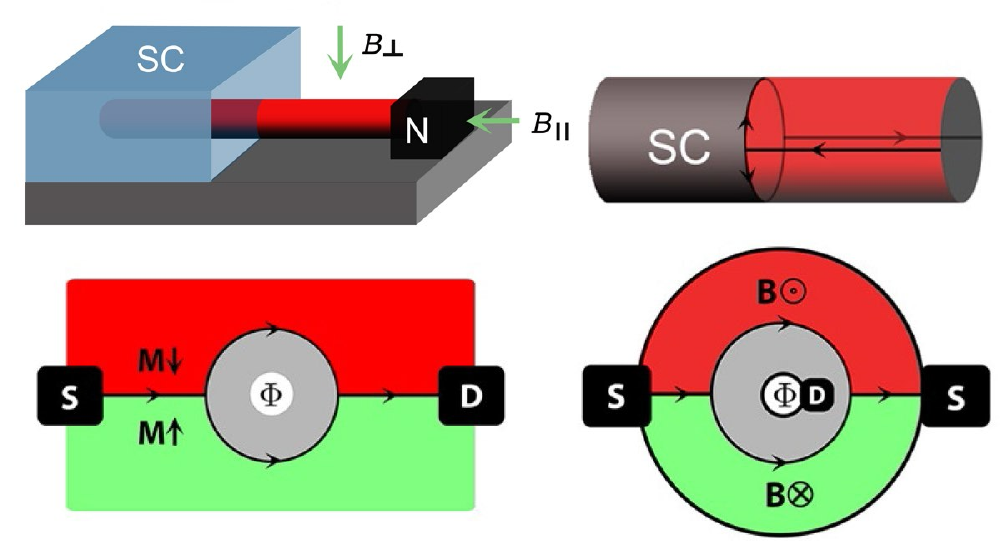}
\llap{\parbox[b]{6.3in}{a)\\\rule{0ex}{1.7in}}}
\llap{\parbox[b]{3.in}{b)\\\rule{0ex}{1.7in}}}
\llap{\parbox[b]{6.5in}{c)\\\rule{0ex}{0.8in}}}
\llap{\parbox[b]{3.1in}{d)\\\rule{0ex}{0.8in}}}
\caption{a) An NS junction formed with a TI nanowire. b) Schematic representation of the modes
involved in transport: a chiral mode splits into two Majorana modes at the interface, recombines,
and exits as a chiral mode again. c) The Majorana interferometer proposed in
Refs.~\onlinecite{FK09,ANB09}. S and D denote source and drain respectively. d) An unfolded
representation of the setup
in b).}\label{setup}
\end{center}
\end{figure}


A prominent example of a 1D topological superconductor is realized in semiconducting quantum wires
in the presence of a magnetic field~\cite{ORO10,LSD10}. Recent transport experiments with such wires
aimed to demonstrate the existence of this phase have reported a finite zero-bias
conductance across a NS junction~\cite{MZF12,DRM12}, but the predicted quantization has so far
remained a challenge to observe. A possible reason is that these wires typically host several
modes~\cite{LSD11,PKR12,LPL12,PDW12,PSA12,RTK13}
and fine tuning the chemical potential to the single mode regime can be difficult. In the presence
of several modes, either a tunnel barrier~\cite{MZF12,DRM12} or a quantum point contact
\cite{WAD11} may be used to isolate the resonant contribution, but the temperature required to
resolve a zero-bias peak then becomes challengingly small. The optimal NS junction to
probe this effect should therefore have a robust, easy to manipulate single-mode normal part
smoothly interfaced with a superconductor that can be controllably driven into the topological
phase. 



In this work, we propose to realize such a junction starting from an alternative route to 1D
topological superconductivity, recently proposed by Cook and Franz~\cite{CF11}, based on
the use of nanowires made from three dimensional topological insulators (TI). In the nanowire
geometry, the 2D surface states of a TI are 
resolved into a discrete set of modes, with the property that when a parallel flux
of $h/2e$ threads the wire, the number of modes is always odd~\cite{ZV10,BBM10,BM13}. When a
superconducting gap is induced on the surface
via the proximity effect, this guarantees that the system becomes a topological
superconductor~\cite{K01,CF11}. An NS junction can then be built be proximitizing only part of the
wire, where the superconducting part can be tuned in and out of the topological phase with
the in-plane flux~\cite{CF11,IBS13}. 

In addition, our design simultaneously allows to drive the normal region into the
single-mode regime by exploiting the unique orbital response of TI surface states to
magnetic fields~\cite{L09,V11,ZWX12,SRA12,BF13}. When a perpendicular field is
applied to the normal part of the wire, its top and bottom regions become insulating
because of the Quantum Hall Effect. In between these regions, counter-propagating chiral edge states
are formed, which are protected from backscattering due to their spatial separation. The resulting
NS junction, shown in Fig.~\ref{setup}, has a single chiral mode reflecting from the superconductor,
and is ideal for probing conductance quantization. Moreover, all of its components are readily
available, as both surface transport in TI nanowires \cite{PLK09,XHW11,HZC13,DVT13} and the
contacting of bulk TI with superconductors~\cite{VSH12,WBG12,CDB13} have already been demonstrated
experimentally. In the remainder of this paper, we provide a detailed
study of the transport properties of this system, demonstrating that conductance quantization is
achievable under realistic conditions, and discuss the advantages of our setup over other
proposals. 





\begin{figure}[t]
\begin{center}
\includegraphics[width=8.6cm]{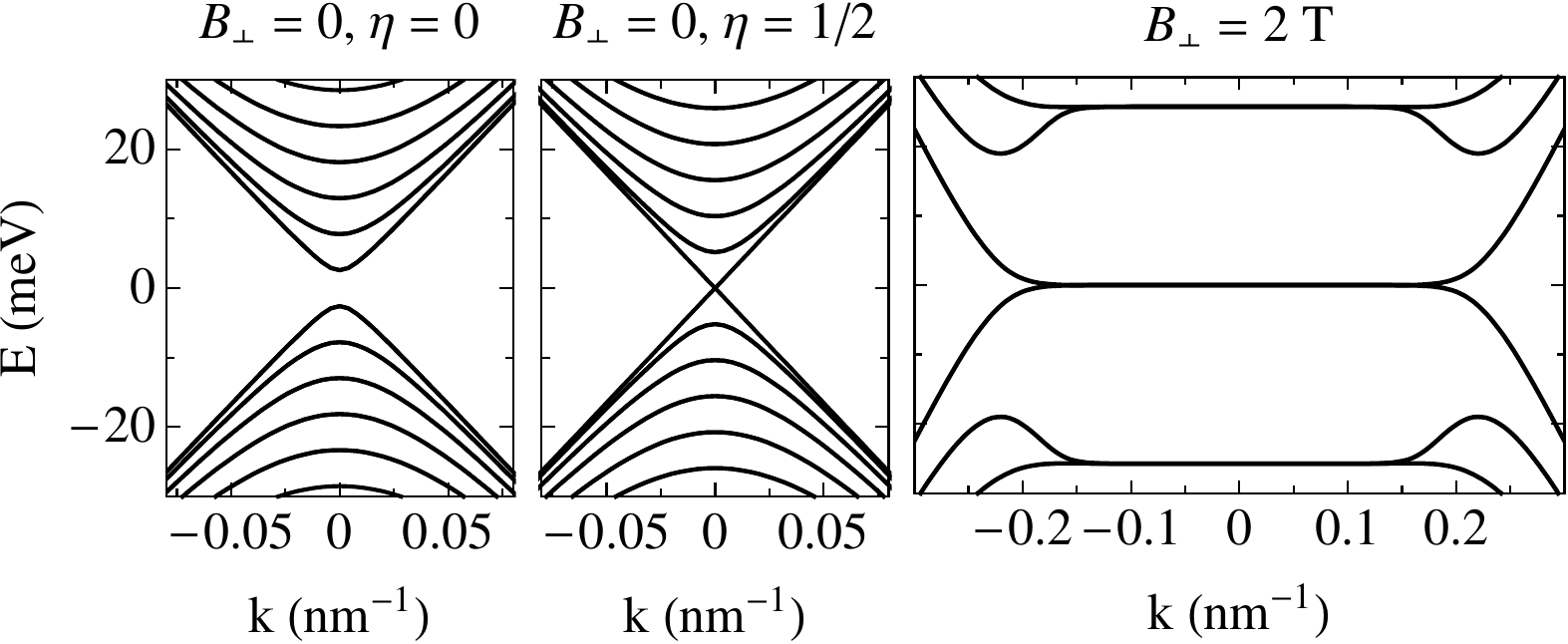}\llap{\parbox[b]{6.65in}{a)\\\rule{0ex}{1.25in}}}
\includegraphics[width=4.25cm]{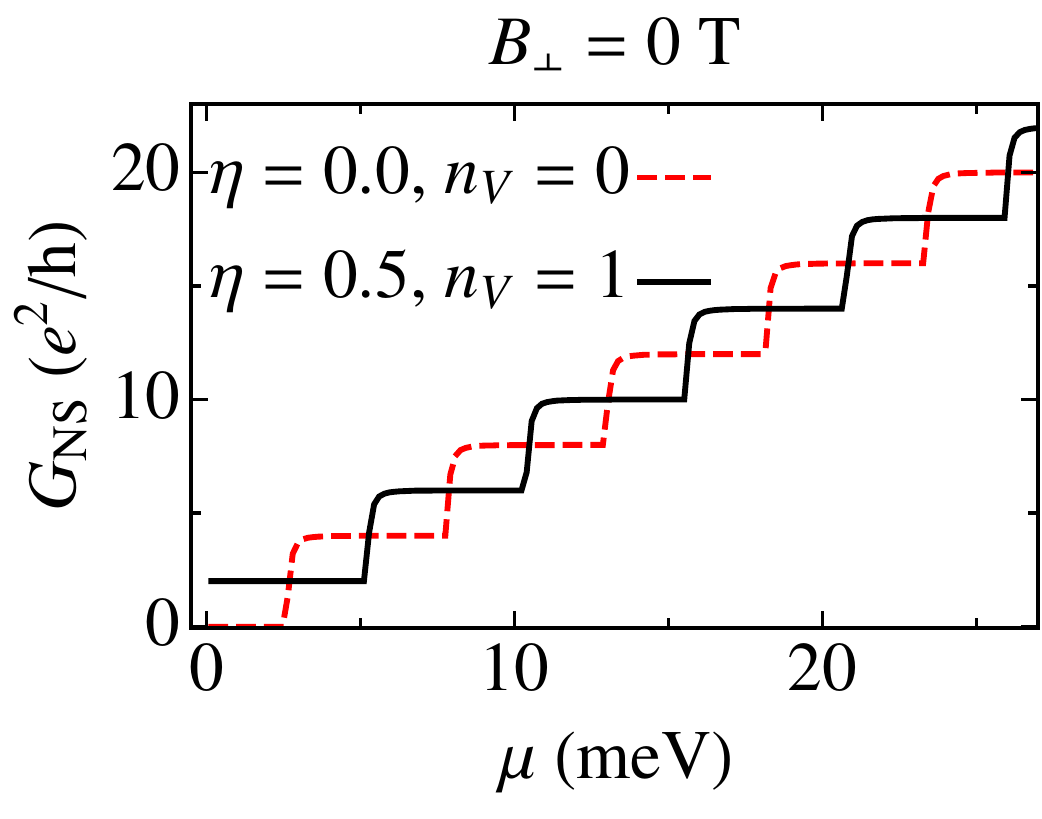}\llap{\parbox[b]{3.2in}{b)\\\rule{0ex}{1.2in}}}
\includegraphics[width=4.25cm]{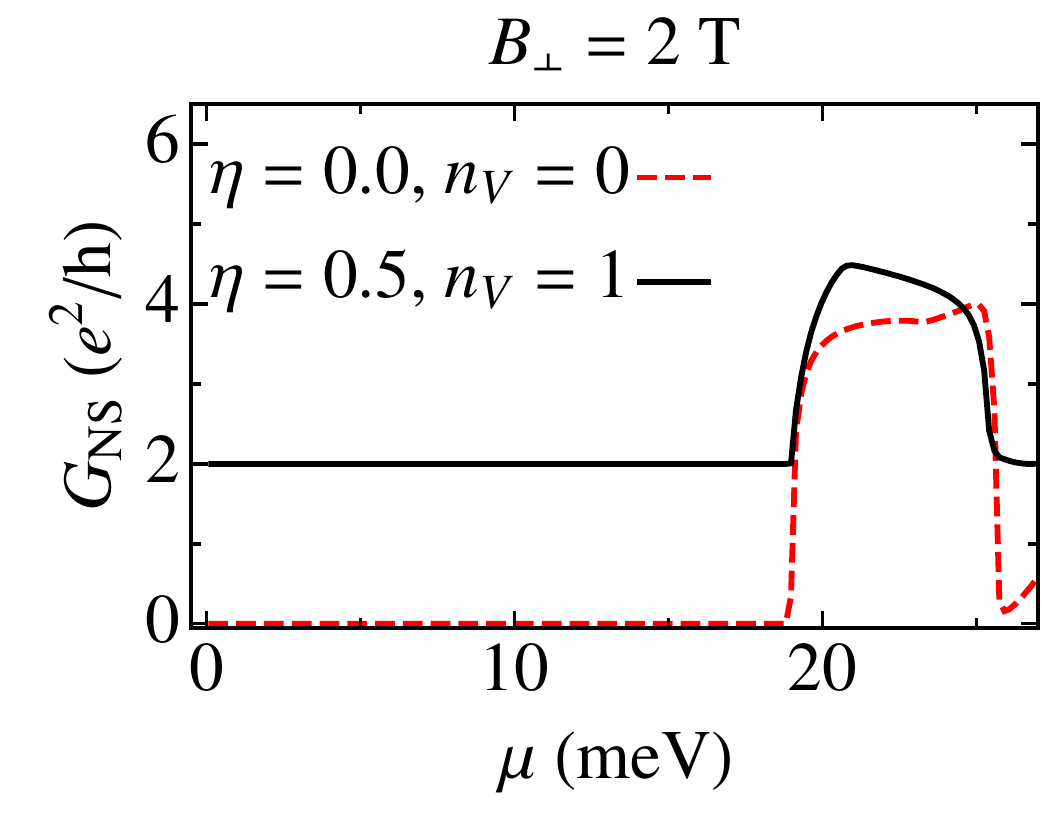}\llap{\parbox[b]{3.in}{c)\\\rule{0ex}{1.2in}}}
\caption{a) The spectrum of a wire of dimensions $h=40$ nm and $w = 160$ nm for $B_{\perp}=0$ and
$\eta=0$ (left), $B_{\perp}=0$ and $\eta=1/2$ (center) and $B_{\perp}=2 T$ (right). Note that in the
last
case the spectrum is independent of $\eta$. b) NS conductance for $B_{\perp}=0$ and $\Delta_0 =$
0.25 meV as a function of $\mu$, for $\eta=0$, $n_v=0$ (dashed line) and $\eta=1/2$,
$n_v=1$ (full line) c) The same for $B_{\perp}=2 T$.}\label{NSG1}
\end{center}
\end{figure}

To model the proposed device, we consider a rectangular TI nanowire of height
$h$ and width $w$ (perimeter $P = 2h+2w$). The surface of the wire is parametrized with two
coordinates $(x,s)$, where $s$ is periodic $s \in [0,2\pi]$ and goes around the perimeter of the
wire, while $x$ goes along its length. We first consider a magnetic field parallel to the wire,
$\vec B = (B_{\parallel},0,0)$, described with the gauge choice $\vec A =
B_{\parallel}(0,-z/2,y/2)$. The dimensionless flux
through the wire is $\eta = B_{\parallel} hw/ (h/e)$. The effective theory
for the surface states is the same as for a cylindrical wire~\cite{BBM10,ZV10}, with $s$ the
azimuthal angle
\begin{equation}
H_0=-iv_F \left[\sigma_x\partial_x + \sigma_y(2\pi/P)(\partial_s +i \eta)\right], 
\end{equation} 
where we set $\hbar=1$ and take $v_F = 330$ $\rm meV\,nm$~\cite{LQZ10}. The wavefunctions satisfy
antiperiodic boundary conditions in $s$ due to the curvature-induced $\pi$
Berry phase \cite{ZV10,BBM10}. The eigenfunctions of $H_0$ thus have the form
\begin{equation}
\psi_{k,n}(x,s) = e^{i k x} e^{i l_n s} \chi_{k,n}, \label{eigs}
\end{equation}
with half-integer angular momentum $l_n = n-1/2$ where $n\in \mathbb{Z}$. The spectrum is
$
E_{k,n} = v_F \sqrt{k^2+(2\pi/P)^2(l_n+\eta)^2},
$
and is depicted in Fig.~\ref{NSG1}(a). For $\eta=0$ all modes are doubly
degenerate, while for $\eta=1/2$ the number of modes is always odd because the $n=0$ one is
not degenerate.

By bringing the wire into contact with an $s$-wave superconductor~\cite{CF11}, as shown in
Fig.~\ref{setup}, an $s$-wave pairing potential $\Delta$ is induced due to the proximity effect. The
Bogoliubov-de Gennes Hamiltonian can be written as $H = \frac{1}{2}
\Psi^{\dagger} \mathcal{H} \Psi$ with
\begin{equation}
\mathcal{H} = \left( 
\begin{array}{cc} 
H_0  & \Delta(s) \\
\Delta^*(s) & - T^{-1}H_0 T
\end{array}\right), \label{BdG}
\end{equation}
where $\Psi
= (\psi_{\uparrow},\psi_{\downarrow},\psi^{\dagger}_{\downarrow},-\psi^{\dagger}_{\uparrow})$ is a
Nambu spinor. The induced pairing potential is $\Delta(s) = \Delta_0 e^{-i n_v s}$,
where the phase of $\Delta$ can wind around the perimeter with vorticity $n_v$. For
$\eta=0$ the ground state has $n_v=0$. Around $\eta=1/2$, however, it should be energetically
favorable for $\Delta$ to develop a vortex~\footnote{This is expected to be stable as the
flux is repelled by the bulk superconductor and can only be trapped in the region occupied by the
wire.}. In an actual experiment, $n_v$ is expected to jump
abruptly as $\eta$ is ramped continuously from zero to $1/2$~\cite{CVF12}. For $\eta$
around 1/2 \emph{and} in the presence of a vortex, the nanowire becomes a topological superconductor
for any $\mu$ within the bulk
gap~\cite{CF11}. 

The presence of the vortex is essential in order to observe
perfect Andreev reflection in our setup
. To see this, consider the Hamiltonian in Eq.~\eqref{BdG} in the
presence of a NS
interface at $x=0$ with $n_v$ vortices. Introducing Pauli matrices
$\tau_i$ acting in Nambu space
\begin{align}
\mathcal{H}^{(n_v)} = &\left[
-i\sigma_x \partial_x +  \sigma_y( -i\partial_s + \eta \;\tau_z)2\pi/P - \mu 
\right]\tau_z \nonumber \\  &+ \Delta_0   \theta(-x) e^{-i \tau_z n_v s}\tau_x . \label{BdGv}
\end{align}

For $n_v=0$, electron states in the normal part have finite angular momentum  $l_n$, see
Eq.~\eqref{eigs}, while hole states have angular momentum $-l_n$, \emph{independently of the
value of} $\eta$. Since angular momentum must be conserved upon reflection, a single
incoming electron can never be reflected as a hole. For $n_v=1$ rotational invariance appears to be
broken by the pairing term, but is explicitly recovered after the gauge transformation
$\Psi \rightarrow e^{i \tau_z s/2} \Psi$, which shifts $\eta\rightarrow \eta-1/2$. This
transformation also changes the boundary conditions to periodic, such that angular momenta
take integer values $l'_n =n$. As a result, the $n=0$ electron state now has the same angular
momentum as its conjugate hole state and can be reflected into it. 

The NS conductance of the junction is computed from the Andreev reflection matrix, evaluated
separately for every $n$, in a very similar way to Ref.~\onlinecite{B06}. To compute
it, we define incoming $\psi^{e-}_n$ and outgoing
$\psi^{e+}_n$ propagating electron states in the normal part, and similarly for hole states
$ \psi^{h-}_n$ and $\psi^{h+}_n$. Normalization is chosen such that all propagating states carry the
same current, $J_x =\left<\psi|\sigma_x |\psi\right> = 1$. These are matched to the
evanescent states in the superconductor $\psi^{S+}_n$ and $\psi^{S-}_n$ by imposing continuity of
the wavefunction at the junction (dropping the label $n$ for ease of notation)
\begin{align}
\psi^{e-} + r_{ee}\psi^{e+} + r_{he} \psi^{h-} &= a\psi^{S+} + b\psi^{S-}, \\
\psi^{h-} + r_{hh}\psi^{h+} + r_{eh} \psi^{e-} &= a' \psi^{S+} + b' \psi^{S-}.
\end{align}
The reflection matrix is defined as
$
r = \left(\begin{smallmatrix}
           r_{ee} & r_{eh} \\
           r_{he} & r_{hh} 
          \end{smallmatrix}\right),
$
and is both unitary and particle-hole symmetric. The conductance is given by
$
G_{\rm NS} = \frac{2e^2}{h} \text{tr} \; r_{eh} r_{eh}^{\dagger},
$
where the trace sums over all propagating modes. The resulting $G_{\rm NS}$ for $\Delta_0 =$ 0.25
meV are shown in Fig.~\ref{NSG1}(b). When $\eta=1/2$, $n_v=1$ and in the range $\mu<\pi/P$, a single
mode is reflected from a topological superconductor resulting in a conductance of $2 e^2/h$.

\begin{figure}[t]
\begin{center}
\includegraphics[height=0.37\columnwidth]{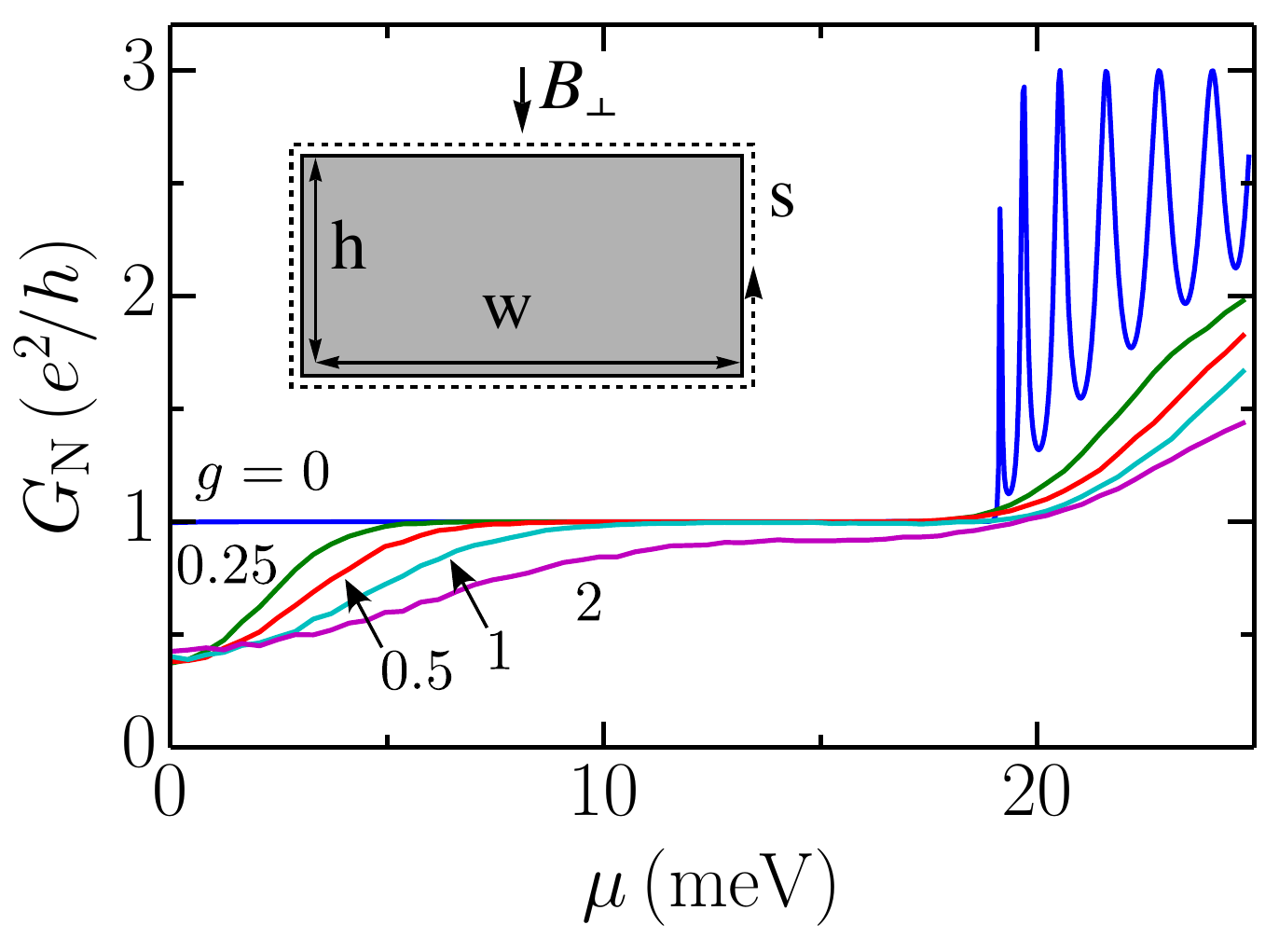}
\includegraphics[height=0.37\columnwidth]{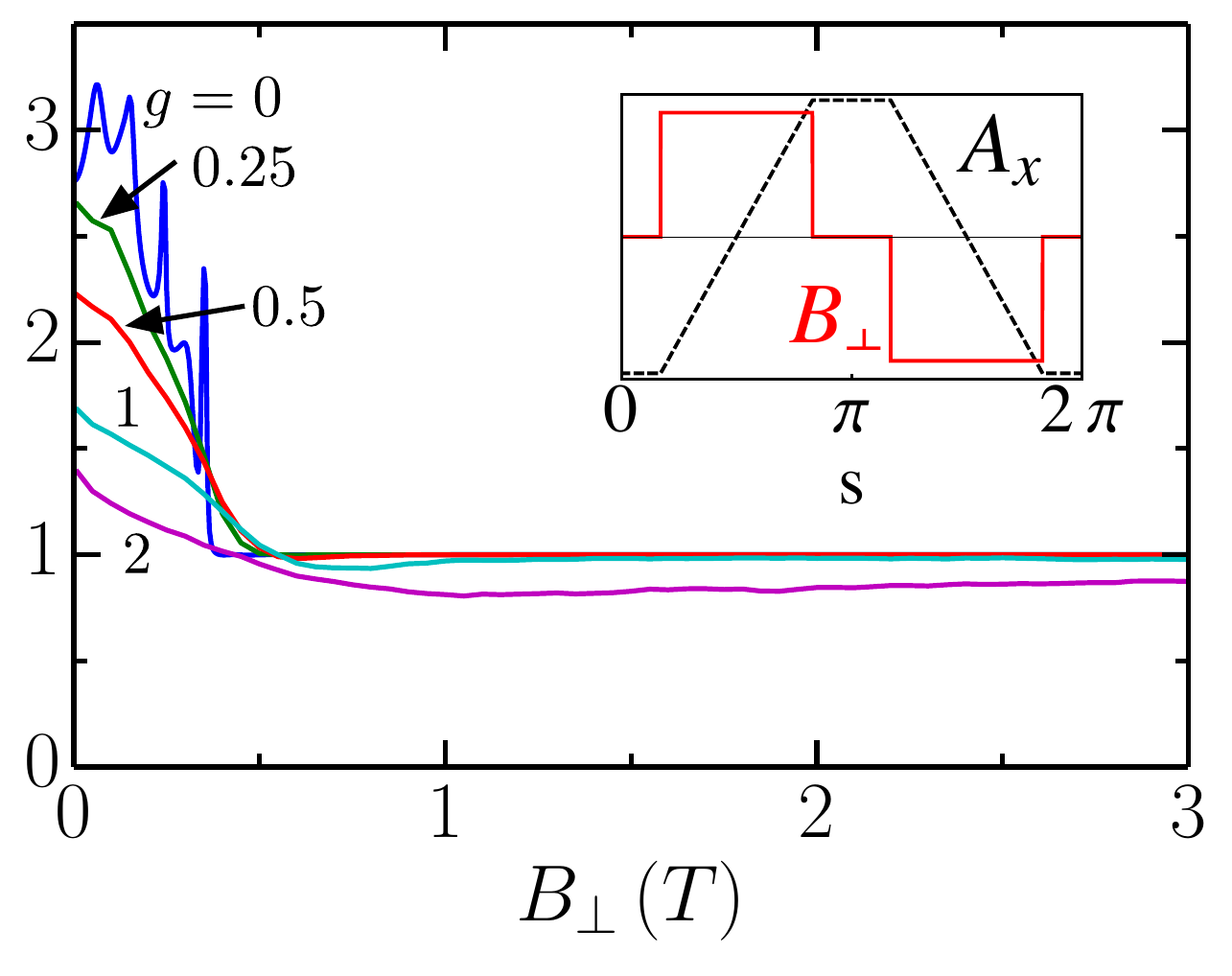}
\caption{Disorder averaged conductance for a finite wire of dimensions $L=400$ nm, $h=40$ nm and $w
= 160$ nm for different values of the disorder strength $g$. a) Conductance as a function of
chemical potential with fixed magnetic field $B_{\perp}=2$ T. Inset: Cross section of the
rectangular wire, with the coordinate $s$ depicted as a dashed arrow. b) Conductance as a function
of $B_{\perp}$ at fixed chemical potential $\mu = 10$ meV. Inset: The vector
potential $A_x$ as given by Eq. \eqref{ax}, and its associated magnetic field
profile.}\label{perpfield}
\end{center}
\end{figure}

The conditions to observe conductance quantization in this setup are not optimal yet, mainly
because the chemical potential has to be tuned into a small gap $\pi/P$. This limitation can be
overcome by the addition of a perpendicular field. Consider the
Hamiltonian of the normal wire with $\vec B = (B_{\parallel},B_{\perp},0)$ and a vector potential
$\vec A = B_{\perp} (z,0,0)+ B_{\parallel}(0,-z/2,y/2)$ such that translational invariance is
still preserved in the $x$ direction
\begin{equation}                      
H=\sigma_x[-i\partial_x + e A_x(s)] + \frac{2\pi}{P} \sigma_y( -i\partial_s + \eta).
\end{equation}
The vector potential in the surface coordinates is
\begin{align}
A_x(s) = B_{\perp} P \left\lbrace \begin{array}{clcl}
-\frac{r}{4} & & \frac{-1+r}{4} & <  \frac{s}{2\pi} < \frac{1-r}{4} \\
\frac{s}{2\pi} - \frac{1}{4}  & & \frac{1-r}{4} & <  \frac{s}{2\pi}  < \frac{1+r}{4} \\
\frac{r}{4} & &\frac{1+r}{4} & < \frac{s}{2\pi}  < \frac{3-r}{4} \\
-\frac{s}{2\pi} + \frac{3}{4} & & \frac{3-r}{4} & <  \frac{s}{2\pi}  < \frac{3+r}{4}
\end{array}\right. ,\label{ax}
\end{align}
with $r = \frac{w}{w+h}$. The profiles of
$A_x$ and $B_{\perp}$ along the $s$ direction are shown in the inset of
Fig.~\ref{perpfield}(b). Since rotational symmetry is broken, the different $n$
modes are mixed. In the angular momentum basis, Eq.~\eqref{eigs}, the Hamiltonian is
$
H = \sum_{n,n'=-N}^N
\chi_{k,n}^{\dagger}H_{n,n'}(k) \chi_{k,n'},
$
where $N$ is an angular momentum cutoff. The matrix element is given by
\begin{align}
H_{nn'}(k) &= \left[\sigma_x k  +\sigma_y (2\pi/P) (n-1/2+\eta )
\right]\delta_{n,n'} \nonumber \\
& +\sigma_x \sum_{m=-M}^M e A^{(m)}_x\delta_{n,n'+m},\label{hperp}
\end{align}
where $A^{(m)}_x = \int_0^{2\pi} \frac{ds}{2\pi} e^{-i m s} A_x(s) = B_{\perp}P
(-1)^{\frac{m+1}{2}}\sin (m \pi r/2) / m^2 \pi^2$ if $m$ is odd and vanishes otherwise, and $M$ is
a cutoff for the number of Fourier components of
$A_x$, with $M\leq N$. The spectrum of the wire only changes qualitatively when $l_B < w$, with $l_B
= (\hbar/eB_{\perp})^{1/2}$ the magnetic length, and Landau levels start to form in the top and
bottom
surfaces, which merge smoothly with dispersing chiral states localized in the sides. 
The spectrum in this regime, shown in Fig.~\ref{NSG1}(a), becomes independent of $B_{\parallel}$.

The NS conductance for finite $B_{\perp}$ can be computed as before
with one important difference: in the basis states for the normal part, the
evanescent states (with Im$[k_{ev}]>0$) must be included to obtain a
well-defined matching condition. The incoming electron states,
labelled now by
$\alpha = 1,\ldots,N_{\rm prop}$, are
$
\psi^{e-}_{\alpha} = e^{-ikx} \sum_{n=-N}^N e^{i l_n s} \chi^{e-}_{n,k},
$
and similarly for $\psi^{e+}_{\alpha}$, $\psi^{h-}_{\alpha}$ and $\psi^{h+}_{\alpha}$. The
evanescent states are defined as $\psi^{N}_{\alpha'}$, with $\alpha' = 1, \ldots,
N_{\rm ev}$, with $N_{\rm ev} + N_{\rm prop} = 2N$. Both propagating and evanescent momenta
and wavefunctions are obtained from the transfer matrix of the normal
part~\cite{LJ81,U97,JIB13}. We
assume
that $B_{\perp}$ is completely screened in the superconducting part of the wire (see
Fig.~\ref{setup}), so that the eigenstates in this region
remain unchanged. Continuity of the wavefunctions at the interface
\begin{align}
&\psi^{e-}_{\alpha} + \sum_{\beta=1}^{N_{\rm prop}}\left[(r_{ee})_{\alpha \beta}\psi^{e+}_{\beta} +
(r_{he})_{\alpha\beta} \psi^{h-}_{\beta}\right] + \\
\sum_{\alpha'=1}^{N_{\rm ev}}&
\left[c_{\alpha \alpha'} \psi^{N,e}_{\alpha'} + d_{\alpha \alpha'}
\psi^{N,h}_{\alpha'}\right] 
= \sum_{n=-N}^N \left[a_{\alpha n} \psi^{S+}_n + b_{\alpha n} \psi^{S-}_n \right].\nonumber
\end{align}
For every value of $\alpha$, we project into angular momentum states with $n = - N,\ldots, N$. 
and since the spinors have four components (spin and particle-hole degrees of freedom) this
yields a system of $8N$ equations with $2N_{\rm prop}+2N_{\rm ev}+4N = 8N$ coefficients. The system
is solved numerically, and the conductance obtained is shown in Fig.~\ref{NSG1}(c). In the
single-mode regime, at zero flux and $n_v=0$ we have $G_{\rm NS}=0$, but at $\eta=1/2$ and $n_v=1$
(when the superconductor is topological), we have $G_{\rm NS}=2 e^2/h$ as expected. 

The quantization of $G_{\rm NS}$ can be understood intuitively in terms of a 1D low-energy model,
depicted in Fig.~\ref{setup}(b), similar to the one describing the Majorana interferometer proposed
in Refs.~\onlinecite{FK09,ANB09} (see also related studies of Majorana
interferometry with chiral Majorana modes~\cite{NA10,STS11,LFB12} and Majorana bound
states~\cite{BP10,F10,HAH10,MPZ11,BS11,LT11,SBW11,JB13,YS14,UY14}). In this model, an incoming
chiral mode leaving the source is
split into two Majorana modes that appear at the interface between the the superconductor and the
regions with finite $B_\perp$~\cite{TZB13}. In the absence of a vortex the two
Majoranas recombine as an electron on the other side of the wire and return to the source through
the channel of opposite chirality, yielding $G_{\rm NS}=0$. However, if a vortex is present, the two
Majoranas accumulate a relative phase of $\pi$ and recombine as a hole, while a Cooper pair is
transferred to the superconductor, yielding $G_{\rm NS}=2e^2/h$.

The quantization of the conductance in our setup is expected to be robust to disorder to some
extent, because transport in the normal part is mediated by spatially separated chiral modes. In
order to test this robustness we introduce disorder into the Hamiltonian  of a normal
wire in the presence of $B_\perp$, and compute the two terminal conductance $G_{\rm N}$ of a finite
size wire numerically, following the method of Ref. \onlinecite{BTB07}. The disorder potential has a
correlator 
$
\langle V(r)V(r')\rangle= g \frac{(\hbar v_F)^2}{2\pi\xi_\text{D}^2}e^{-|r-r'|^2/2\xi_\text{D}^2},
$
with $\xi_\text{D}$ the disorder correlation length and $g$ a dimensionless measure of the
disorder strength. Our data is obtained by averaging over $10^{3}$ disorder
configurations. The results are shown in Fig.~\ref{perpfield}. In the
single-mode regime, the conductance of the normal wire indeed remains quantized to $e^2/h$ in the
presence of
moderate disorder, as long as the chemical potential is not very close to zero. The conductance for
each disorder realization is also quantized. A full
characterization of the effects of disorder will be presented in a future work~\cite{JIB13}.

\emph{Discussion} -  An important feature of our proposal is that all effects induced
by the
magnetic field are of purely orbital origin. The Zeeman coupling will be a small correction at the
fields considered, and does not change our predictions qualitatively~\cite{JIB13}. In our setup, a
quantized conductance can be obtained with both $B_\perp=0$ and
finite $B_{\perp}$, but the latter case has several advantages that are worth stressing. First, the
single-mode regime remains accessible for chemical potentials
ranging up to values of the
order of the cyclotron frequency $\omega_c$, rather than the finite size gap $\pi/P$. Second,
chiral mode transport in the normal part is robust against
finite disorder due to spatial separation of counter-propagating chiral modes.  Third, the spectrum
of the normal part in the presence of $B_{\perp}$ becomes independent of $B_{\parallel}$, which
 affects only the superconducting part. $B_{\parallel}$ thus becomes an independent knob
driving
the transition from a trivial to a topological superconductor, while the chiral modes remain intact.
In this case, measuring $G_{\rm NS}=0$ would represent a genuine consequence of reflection from a
trivial superconductor, as opposed to the $B_{\perp}=0$ case where this value of $G_{\rm NS}$ could
result from an insulating normal part, see Fig.~\ref{NSG1}(b). 

Our proposal realizes a version of the Majorana interferometer with some important
differences. In our setup, instead of contacting the two chiral modes separately the source
electrode contacts both channels and the superconductor is the
drain~\cite{A12}, see Figs.~\ref{setup}(c-d). In addition, the original proposals use ferromagnets
and a finite superconducting island to create the Majorana modes, while our setup uses a bulk
superconductor and a homogeneous magnetic field~\cite{TZB13}, making it experimentally more
feasible. Despite these differences, the finite voltage and finite temperature behavior of $G_{\rm
NS}$will be similar to those in Refs.~\onlinecite{FK09,ANB09}. This introduces an important
advantage to our setup over current semiconducting wires, where the temperatures required to observe
conductance quantization are of the order of mK. In our setup, the limiting
temperature is determined by the proximity induced gap~\cite{ANB09}. Assuming 
$\Delta_0\approx0.1-0.25$ meV~\cite{MZF12,DRM12,DYH12} this corresponds to 1-3 K.

Finally, we note that screening $B_{\perp}$ in the SC region requires the use of a superconductor
with a high critical field. For example, the superconductor could be a Ti/Nb/Ti trilayer as the one
used in the experiment in Ref.~\onlinecite{DYH12}, which was estimated to have
$H_{c1}= \rm
2.5T$. 



We thank J. Dahlhaus, J.E. Moore, A. Vishwanath and J. Analytis for useful discussions. We
acknowledge financial support from the ``Programa Nacional de Movilidad de Recursos Humanos"
(Spanish MECD) (F. de J.) and DARPA FENA (R.I and J.H.B.). R. I. is an Awardee of the Weizmann
Institute of Science National Postdoctoral Award Program for Advancing Women in Science.

\bibliography{wires}






\end{document}